# Apparent extracellular current density and extracellular space: basis for the current source density analysis in neural tissue

**Hiroyoshi Miyakawa**[(1)] and **Toru Aonishi**[(2)].

([(1)]Laboratory of Cellular Neurobiology, School of Life Sciences, Tokyo University of Pharmacy and Life Sciences, Tokyo, Japan ; [(2)]Department of Computational Intelligence and Systems Science, Tokyo Institute of Technology, Yokohama, Japan )

## Abstract

This article provides a theoretical basis for relating macroscopic electrical signals recorded from biological tissue, such as electroencephalogram (EEG) and local field potential (LFP), to the electrophysiological processes at the cellular level in a manner consistent with Maxwell's equations. Concepts of the apparent extracellular current density and the apparent extracellular space with apparent permittivity and conductivity are introduced from the conservation of current and Gauss's theorem. A general equation for the current source density (CSD) analysis is derived for biological tissue with frequency-dependent apparent permittivity and conductivity. An intuitive account of the apparent extracellular space is given to relate the concept to the dielectric dispersion of biological tissue.

21 pages, 5 figures

## Contents

# 1. Introduction

Important aspects of neural function are made possible by electrical activity of neurons, and therefore various kinds of electrical signals are recorded to monitor activity of neural systems. Those signals include membrane potential, membrane current, local field potential (LFP), unit recording, electro-encephalogram (EEG), magneto-encephalogram (MEG), Electro-corticogram (ECog), and so on. Theories to describe electrical activity of biological cells have been developed in a research field called electrophysiology and form the basis for neuroscience. Unfortunately, however, it is not obvious how the electrophysiological descriptions of electricity in biological systems are related to the electrodynamical theory, more precisely Maxwell's equations (Maxwell, 1873). Although it may not be a problem if one is interested in activity at the cellular or membrane level, this situation causes unnecessary confusion when one attempts to relate macroscopic signals, such as EEG or LFP, to electrical activity of individual neurons or parts of neurons. Here is one example. The equation below describes the relation between the extracellular potential $\phi$ and the membrane potential $I_m$ (here $\sigma$ denotes conductivity), and has been used in the current source density (CSD) analysis (Nicholson & Llinas, 1971; Nicholson, 1973; Nicholson & Freeman, 1975; Mitzdorf, 1985; Pettersen et al., 2006; Bedard & Destexhe 2011).

$$\sigma \nabla^2 \phi = -I_m \qquad (1)$$

This equation has a form similar to Poisson's equation that describes the relation between the electrical potential $\phi$ and the charge density $\rho$ (here $\varepsilon$ denotes permittivity).

$$\nabla^2 \phi = -\frac{\rho}{\varepsilon} \qquad (2)$$

This kind of similarity seems to give the false impression that the eq. (1) can be readily derived from the basic classical electrodynamics. One might misunderstand that electric fields in the brain can be understood if only one understands Maxwell's equations. It is not true. Despite the mathematical similarity, equations (1) and (2) describe quite different situations. To derive equation (1), one needs to realize that a neural tissue is an ensemble of many cells, and that the cell membrane separates the tissue into two parts, extracellular and intracellular space. As will be shown in this article, applying Gauss's divergence theorem to current gives rise to equation (1). By contrast, Poisson's equation is derived from Gauss's law that describes the relation between the electric field and charge (not the current). Such confusion seems to hinder interplay of the cellular level electrophysiology and the macroscopic brain science. We should properly understand how the extracellular potential and current are related to the membrane current.

There is another matter one needs to consider in relating electrophysiology with macroscopic electrical signals. It is the marked dielectric dispersion in biological tissue. In both microscopic electrophysiology and macroscopic brain science, it is usually assumed that the capacitive current in the extracellular space is negligibly small compared to the conductive current (Nunez & Srinivasan, 2006). But, it has been acknowledged for a long time that the relative permittivity of biological tissue can be as high as $10^6$ in the frequency range lower than 100 Hz (Cole & Cole, 1941). A recent study reported even higher value (Gabriel et al., 1996). Because the frequency range the neuroscientists are interested most is the range lower than 100 Hz, we should not hastily ignore extracellular capacitive current. We should have a theoretical framework with which EEG and LFP can be analyzed without ignoring the capacitive components.

In this article, we derive a concept of the "apparent extracellular current" and a general form of the current source density analysis from Maxwell's equations in Section 3. Before doing so, we give an overview of the classical electrodynamics in dielectric material media in Section 2 in order to avoid confusion of the terms such as total current and free charge. On the basis of the "apparent extracellular current", we provide an intuitive account for the concept of "apparent extracellular space" in Section 4. This article does not consider electric fields generated by varying magnetic field, in other words, it assumes that the quasi-static condition holds in neural tissue.

## 2. Overview of Quasi-Static Electrodynamics in Material Media

Our goal is to understand how extracellular potential and current in biological tissue are related to the membrane current in a manner consistent with the classical electrodynamics. Therefore, we would like to start by giving an overview of electrodynamics in dielectric material media, and restrict our attention to the spatial scale greater than 1 micron and the frequency range slower than 10 kHz.

### 2-1. Fundamental electrodynamics in continuous media
*Maxwell's equations*

Electrodynamics in dielectric material media is described by **Maxwell's equations** which are a set of following four equations.

$$\nabla \bullet D = \rho_{free} \qquad \text{Gauss's law} \qquad (3)$$

$$\nabla \times E = -\partial B / \partial t \qquad \text{Faraday's law} \qquad (4)$$

$$\nabla \bullet B = 0 \qquad \text{non-existence of magnetic monopoles} \qquad (5)$$

$$\nabla \times H = J_{free} + \partial D / \partial t \qquad \text{Ampere's law modified by Maxwell} \qquad (6)$$

Here, $E$ is the **electric field**, and $B$ is the **magnetic flux density** (or **magnetic induction**). The **electric flux density** (or **electric displacement**) $D$ and the **magnetic field** $H$ are defined as follows.

$$D = \varepsilon_0 E + P \qquad (7)$$

$$H = \frac{1}{\mu_0} B - M \qquad (8)$$

Here, $P$ is the **electric polarization**, $M$ is the **magnetic polarization**, $\varepsilon_0$ is the **permittivity** or **dielectric constant of free space** ($\varepsilon_0 = 8.85 \times 10^{-12}$ F/m) and $\mu_0$ is the **magnetic permeability of free space**. The **free charge** (the density to be precise) $\rho_{free}$ that appears in Gauss's law arises from the separation of all the charge into those due to the electric polarization $\rho_{polarization}$ and the rest.

$$\rho_{all} = \rho_{free} + \rho_{polarization} \qquad (9)$$

The (density of the) **all charge** in the system $\rho_{all}$, and the (density of the) **polarization charge** $\rho_{polarization}$ are defined as follows (Feynman, 1963).

$$\nabla \bullet \boldsymbol{E} = \rho_{all}/\varepsilon_0 \tag{10}$$

$$\rho_{polarization} = -\nabla \bullet \boldsymbol{P} \tag{11}$$

The free charge density $\rho_{free}$ is charge density excluding the charge due to polarization. The term $\boldsymbol{J}_{free}$ that appears in the modified Ampere's law, eq. (16), is the **free charge current** (the current density to be precise) carried by $\rho_{free}$, the free charge (density).

*Conservation of current*

The time derivative of the electric flux density (or the displacement) $\partial \boldsymbol{D}/\partial t$ is called the **displacement current** (density) $\boldsymbol{J}_{displacement}$. In this article, we refer to the summation of the free charge current density and the displacement current as the **total current** $\boldsymbol{J}_{total}$.

$$\boldsymbol{J}_{total} = \boldsymbol{J}_{free} + \boldsymbol{J}_{displacementl} \tag{12}$$

$$= \boldsymbol{J}_{free} + \frac{\partial \boldsymbol{D}}{\partial t} \tag{13}$$

$$= \boldsymbol{J}_{free} + \frac{\partial (\varepsilon_0 \boldsymbol{E} + \boldsymbol{P})}{\partial t} \tag{14}$$

$$\boldsymbol{J}_{total} = \boldsymbol{J}_{convection} + \boldsymbol{J}_{conduction} + \boldsymbol{J}_{polarization} + \partial(\varepsilon_0 \boldsymbol{E})/\partial t \tag{15}$$

The free charge current $\boldsymbol{J}_{free}$ can either be the **convection current** $\boldsymbol{J}_{convection}$ (current results from the motion of charge such as electrons in a vacuum tube) or the **conduction current** $\boldsymbol{J}_{conduction}$ (current results from the drift of charge under the influence of electric field such as ions in electrolytes). The displacement current $\boldsymbol{J}_{displacement}$ is a summation of $\partial \boldsymbol{P}/\partial t$ and $\partial(\varepsilon_0 \boldsymbol{E})/\partial t$. The term $\partial \boldsymbol{P}/\partial t$ is the **polarization current** $\boldsymbol{J}_{polarization}$ results from the displacement of matter that causes polarization. The other term $\partial(\varepsilon_0 \boldsymbol{E})/\partial t$ is the displacement current in vacuum, which results from the distortion of the vacuum space. (In some text books (e.g. Griffiths, 1998), the term "displacement current" is used as the name for $\partial(\varepsilon_0 \boldsymbol{E})/\partial t$. In this article we save the name displacement current for $\partial \boldsymbol{D}/\partial t$ since we consider it most appropriate.) With this classification of current, a part of current due to a movement of charge with physical entity in a material will be included in the displacement current, and another part will be included in the conduction current (Feynman 1963).

There is one more important relation implicitly involved in Maxwell's equations. By calculating the divergence of both the left- and right-hand side of Ampere's law, we obtain the following.

$$\nabla \bullet \left( \boldsymbol{J}_{free} + \frac{\partial \boldsymbol{D}}{\partial t} \right) = \nabla \bullet \boldsymbol{J}_{total} = 0 \tag{16}$$

This states the **conservation of total current** $\boldsymbol{J}_{total}$. This holds as long as Maxwell's equations hold. By substituting Gauss's law into eq. (16), we obtain the following.

$$\nabla \bullet \boldsymbol{J}_{free} = -\partial \rho_{free}/\partial t \tag{17}$$

This states the **conservation of charge**, in other words, the **continuity of current.**

*constituent relations: permittivity ε and conductivity σ*

It is empirically known that the electric polarization $P$ and magnetic polarization $M$ are function of electric field and magnetic field, respectively, and hence the flux densities $D$ and $B$ can be related to fields $E$ and $H$. These empirical relations are referred to as the **constituent relations**. In wide range of situations, it is sufficiently accurate to assume that $D$ and $B$ are linearly related to $E$ and $H$, respectively.

$$D = \varepsilon E \tag{18}$$
$$B = \mu H \tag{19}$$

Here a scalar value $\varepsilon$ is called the **permittivity or dielectric constant**, and $\mu$ the **magnetic permeability.** Those media in which linear relations (18) and (19) hold are referred to as **linear media**.

With regard to the conduction current (density) $J_{conduction}$, it is empirically known that the current can be linearly related to the electric field in many situations, and the relation is referred to as **Ohm's law**.

$$J_{conduction} = \sigma \cdot E \tag{20}$$

Here $\sigma$ is referred to as the **electric conductivity**.

*Quasi-static condition*

In biological tissues, the electromagnetic induction is negligiblly small in the frequency range lower than 1 MHz (Nunes & Srinivasan, 2006). Therefore the right-hand side of the eq. (4) can be assumed to be zero.

$$\nabla \times E = 0 \tag{21}$$

In this situation, there are no need to take $H$ and $B$ into consideration if one is interested only in the electirc current and field. The following two equations, Gauss's law and the conservation of current, are all that one needs to solve.

$$\nabla \bullet D = \rho_{free} \tag{3}$$
$$\nabla \bullet \left( J_{free} + \frac{\partial D}{\partial t} \right) = 0 \tag{16}$$

Such condition is referred to as the **quasi-static condition.** When eq. (21) holds, a scalar potential referred to as the **electric potential** $\phi$ can be defined such that the electric field $E$ is the negative gradient of $\phi$.

$$E = -\nabla \phi \tag{22}$$

By substituting (22) into Gauss's law, we obtain Poisson's equation (23).

$$\nabla^2 \phi = -\frac{\rho_{free}}{\varepsilon} \tag{23}$$

When the charge density $\rho_{free}$ is zero, the right-hand side of eq. (23) is zero and this equation is reffered to as Laplace's equation.

$$\nabla^2 \phi = 0 \tag{24}$$

## 2-2. Classification of current and charge in dielectrics

Before applying eq. (3) and (16) to biological tissue, we think it necessary to give more thoughts to the free charge current $J_{free}$, free charge $\rho_{free}$ and the displacement current $\partial D/\partial t$. What exactly are the free charge and free charge current that appear in eq. (3) and (16)?  As an example of a dielectric material, let us imagine a rod-shaped agar gel. When one considers a piece of agar gel as a dielectric material, there is no need to have knowledge of the precise density of the electrons or atomic nuclei, not even the precise density of anions and cations. Anywhere in the rod, the average charge density is zero because the density of the positive and negative charge is the same everywhere. Now imagine that we apply a current from one end of the rod to the other. Then after a while, there will be a gradient of electric potential along the longitudinal axis of the rod. The positive and negative charge we provide to the ends of the ager gel rod by applying the current creates the primary electric field, and the field displaces the electrons and nuclei to create polarization charge. The summation of the primary electric field and the field due to the polarization charge determine the gradient of electric potential in the agar rod. At this point of time, there are same amount of positive and negative polarization charge created within the rod; the positive charge is located near one end and the negative charge near the other end. The total current that has flowed within the rod up to this point of time is the summation of the movement of electrons and nuclei in addition to the current due to distortion of the vacuum space.

Some part of the total current is considered to be the conduction current due to the movement of free charge, but there is an ambiguity regarding which part is considered to be the conduction current. Electric polarization involves various kinds of processes from the electronic polarization with very fast relaxation time to polarization due to slow drift of charged macromolecules with slow relaxation time. We can classify total polarization to fast- and slow-polarization on the basis of relaxation time. The charge generated by the fast-polarization are referred to as the polarization charge of the fast-polarization, and the current due to fast-polarizations contribute to the displace current or capacitive current. The so called free charge is the rest of the charge "other" than those generated by the fast-polarization (Feynman, 1963). This "other" charge is the charge that conveys conduction current that is usually related to the electric field with Ohm's law. In case of the agar-gel rod, one might consider relaxation processes slower than one micro second, such as the electronic polarization and the orientation polarization, as "fast-polarization"; then slow-polarization due to the displacement of anions and cations can be considered to be the conduction of free charge. Here, a large number of anions and cations in the agar provide the basis for the free charge; but, it does not imply that they are the free charge.  Although the center of the gravity of anions and cations drift very slowly, if one pays attention only to the positive and negative polarization charge that appear at the opposite ends of the rod as results of the displacement of anions and cations, one can conceive that some small amount of positive and negative charge move very quickly from one end to the other in opposite directions. These movements constitute what we call the conduction current. Therefore, the free charge in this case is the small amount of imaginary charge that is considered to flow within the rod to generate the slow-polarization charge, not the large number of anions and cations per se. The similar consideration applies to the conduction current and free electrons in metallic conductors. (Feynman、1963; Jackson, 1998; Fano et al., 1960) .

To understand what was stated above, it would help to describe the concepts in mathematical forms. Let us start by dividing polarization *P* to fast and slow components.

$$P = P_{slow} + P_{fast} \tag{25}$$

Substituting this into eq. (14), the total current can be written as follows.

$$J_{total} = J_{convection} + \left(\partial P_{slow}/\partial t\right) + \left(\partial P_{fast}/\partial t\right) + \partial(\varepsilon_0 E)/\partial t \tag{26}$$

The time derivative of $P_{slow}$ is the conduction current $J_{conduction}$ (or $J_{slow\text{-}polarization}$). The time derivative of $P_{fast}$ is the polarization current $J_{polarization}$ (or $J_{fast\text{-}polarization}$ to be more precise).

$$J_{total} = J_{convection} + J_{conduction} + J_{fast-polarization} + \partial(\varepsilon_0 E)/\partial t \tag{27}$$

The first two terms on the right-hand side constitute the free charge current $J_{free}$. Because there is no need to consider $J_{convection}$ in material media, $J_{free}$ consists of the conduction current $J_{conduction}$ in material media. The current $J_{conduction}$ can be referred to as resistive current $J_{resistive}$ or ohmic current.

$$J_{free} = J_{convection} + J_{conduction} = J_{conduction} = \left(\partial P_{slow}/\partial t\right) \tag{28}$$

The last two terms on the right-hand side of eq. (26) and (27) constitute the displacement current $J_{displacement}$, i.e. $\partial D/\partial t$. It can also be called capacitive current $J_{capacitive}$.

$$J_{displacement} = J_{capacitive} = J_{fast-polarization} + \partial(\varepsilon_0 E)/\partial t = \left(\partial P_{fast}/\partial t\right) + \partial(\varepsilon_0 E)/\partial t \tag{29}$$

Note that the currents mentioned here are current densities to be precise. Using these classifications, the conservation of the total current can be written as follows.

$$\nabla \bullet \left(J_{free} + \frac{\partial D}{\partial t}\right) = \nabla \bullet \left(J_{resistive} + J_{capacitive}\right) = 0 \tag{30}$$

Similarly, charge can be divided and classified as follows.

$$\rho_{all} = \rho_{convection} + \rho_{slow-polarization} + \rho_{fast-polarization} \tag{31}$$

The first two terms on the right-hand side constitute the free charge density $\rho_{free}$, and the current due to $\rho_{free}$ is the free charge current $J_{free}$. Because there is no need to consider $J_{convection}$ in material media, $\rho_{free}$ represents the slow polarization charge $\rho_{slow\text{-}polarization}$ in material media. Conductive current is what gives rise to slow polarization charge $\rho_{slow\text{-}polarization}$.

$$\rho_{free} = \rho_{convection} + \rho_{slow-polarization} = \rho_{slow-polarization} \tag{32}$$

Capacitive current is the summation of charge movements that gives rise to fast-polarization charge $\rho_{fast\text{-}polarization}$ and the current $\partial(\varepsilon_0 E)/\partial t$. There is no charge that conveys the current $\partial(\varepsilon_0 E)/\partial t$ because this current is generated by the distortion of the vacuum field, not by the movement of charge with physical entity. Now it must be clear that the fast- and slow-polarization charge are all the charge that constitute $\rho_{all}$ in case of material media. Neither the entirety of all the anions and cations, nor the entirety of all the electrons and nuclei are what constitute $\rho_{all}$. Equation (10) defines $\rho_{all}$ as the charges that gives rise to the electric field $E$. By using these classifications, the conservation of charge can be written as follows.

$$\nabla \bullet J_{free} = -\partial \rho_{free}/\partial t \tag{17}$$

$$\nabla \bullet J_{conduction} = \left(\partial \nabla \bullet P_{slow}/\partial t\right) = -\partial \rho_{slow-polarization}/\partial t \tag{33}$$

Equation (33) implies that "the generation of the slow-polarization charge (right-hand side) is considered to be the result of the conduction of current (left-hand side) in material media".

Although both the eq. (16), the conservation of the total current, and the eq. (17), the conservation of charge, are correct and useful, one needs to understand the meanings of the free charge and the free charge current in applying them.

## 2-3. Electric fields in dielectrics with frequency-dependent permittivity and conductivity

Generally the permittivity and conductivity depend on the location and time. Here, we would like to derive a general equation for the electric field and electric potential in such situations. To do so, we express the electric flux density $D(x,t)$ and the free charge current $J_{free}(x,t)$ as the following (Landau, 1982; Jackson, 1998; Bedard & Destexhe, 2009).

$$D(x,t) = \int \varepsilon(x,\tau)E(x,t-\tau)d\tau \tag{34}$$

$$J_{free}(x,t) = \int \sigma(x,\tau)E(x,t-\tau)d\tau \tag{35}$$

Here, $\varepsilon(x,t)$ and $\sigma(x,t)$ respectively denotes the permittivity and conductivity that depends on location and time. The reason to use such expressions is as follows. In material media, an electric field $E$ would induce polarization, which we refer to as the **dielectric polarization**. A polarization can be considered to be a relaxation process induced by the field. The relaxation cannot be induced instantaneously; it requires some relaxation time to take place. In a matter one works on, the polarization might involve various kinds of relaxation processes with various relaxation times, some are fast and some are slow. Therefore, the polarization at certain point in time should depend on the history of the electric field and the relaxation processes induced by the electric field. The resultant electric flux density $D(x,t)$ should be a convolution of the effects, some are fast and some are slow, induced by the electric field of the past time. If the effects are **linear** to the electric field $E$, we can use eq. (34) to describe the electric flux density $D(x,t)$.

Similar reasoning applies to eq. (35). Because there is no need to consider the convection current in material media, only the conduction current needs to be considered. In many situations, Ohm's law is valid to describe the conduction current. But, there must be very short period of time before the slow movement of charge responsible for slow-polarization is accelerated by the force due to electric field and reach a situation in which no more acceleration is possible. This is the reason that the convolution of slow-polarization due to the field $E$ in the past is necessary to describe the conduction current.

By combining eq. (34) and (35) with the conservation of current, we obtain conditions for the electric field $E$ in a material media with time- and location-dependent permittivity and conductivity. To do so, we use Fourier transforms.

$$E(x,t) = \int E(x,\omega)e^{i\omega t}d\omega \tag{36}$$

$$D(x,t) = \int D(x,\omega)e^{i\omega t}d\omega \tag{37}$$

$$J_{free}(x,t) = \int J_{free}(x,\omega)e^{i\omega t}d\omega \tag{38}$$

By substituting these formulas (36), (37) and (38) into eq. (16), we easily obtain the following.

$$\nabla \cdot (\sigma(x,\omega) + i\omega\varepsilon(x,\omega))E(x,\omega) = 0 \tag{39}$$

This is the general equation for the Fourier transform of the electric field. Under the quasi-static conditions, the electric potential $\phi(x,t)$ can be defined as follows.

$$E(x,t) = -\nabla \phi(x,t) \tag{40}$$

Then the Fourier transform of the electric potential should satisfy the following.

$$\nabla \cdot (\sigma(x,\omega) + i\omega\varepsilon(x,\omega))(-\nabla \phi(x,\omega)) = 0 \tag{41}$$

Here $\phi(x,\omega)$ is the Fourier transform of $\phi(x,t)$.

$$\phi(x,t) = \int \phi(x,\omega) e^{i\omega t} d\omega \tag{42}$$

If the media is isotropic, which implies that both $\sigma$ and $\varepsilon$ are independent of the location, then the equation can be simplified as follows.

$$\nabla^2 \phi(x,\omega) = 0 \tag{43}$$

## 3. Mesoscopic scale Electrodynamics in Biological Tissue : The CSD analysis

### 3-1. Conservation of current in biological tissue

In this part, we will derive equations that describe mesoscopic scale electrodynamics in biological tissue. Despite the fact that biological tissue is quite inhomogeneous, we consider the biological tissue as fairly homogenous dielectric media when we analyze extracellular signals such as EEG and LFP. In estimating current sources for EEG or LFP signals, we use the concept of the conductivity and permittivity of tissue. To properly understand electrodynamics in biological tissue, we need to define the **apparent extracellular current**, **apparent conductivity and permittivity** of the tissue. By properly defining these, we obtain a general form of the current source density analysis. To do so, we would like to start by considering that biological tissue is composed of the interstitial space, the biological membrane, and the intracellular space, because the spatial scale we are interested ranges from few microns to several centimeters. In this range, the cell membrane is the most important factor determining the electrodynamical properties of the tissue because of the very low resistivity and very high capacitance.

Here we use three kinds of tools to derive the concepts of the apparent extracellular current, apparent conductivity and permittivity of biological tissue. Firstly, the conservation of total current, secondly Gauss's (divergence) theorem, and thirdly the coarse graining.

① <u>Conservation of total current</u>

$$\nabla \bullet \left( J_{free} + \frac{\partial D}{\partial t} \right) = \nabla \bullet J_{total} = 0 \tag{16}$$

② <u>Gauss's (divergence) theorem</u>

Gauss's theorem states that "the surface integral of the flux of a vector field $A$ over a closed surface $S$ is equal to the volume integral of the divergence of $A$ over the volume enclosed by the surface $S$".

$$\iiint_V \nabla \bullet A \, dv = \iint_S A \cdot n \, ds \tag{44}$$

(In eq. (44), $n$ denotes a unit vector perpendicular to the surface $S$.)

③ <u>Coarse graining</u>

Integrate and average a current density defined in a small volume of tissue over the volume, shrink the volume to an infinitesimally small volume element, and assign the averaged density to this infinitesimal volume element.

### *Gauss's theorem in a mesoscale volume that includes cell elements*

Consider a volume $V$ enclosed by a closed surface $S$ in a biological tissue; for example a cube of side length 10 micron. A part of the volume may be an intracellular space, and some other part may be an interstitial space. Some cells may be totally contained, and some cell may be only partly contained in this volume. For this volume, we will define the "apparent extracellular current density". To prepare for this, we need to divide the volume and surface to smaller parts and name them. We apply the names and symbols used by Nicholson (1973) with some modifications (Figure 1). The cellular parts are called cores. Let the volume contain $m$ cores each having a surface $M_i$ and volume $O_i$ within $S$, and let these cores intersect the surface $S$ at $n$ disks $N_j$. Let the total summation of $N_j$ be $S'_{int}$, and let the surface $S$, less the disks $S'_{int}$, be $S'_{ext}$. Let the total summation of $O_i$ be $V'_{int}$, and let the volume $V$, less the volume $V'_{int}$ be $V'_{ext}$. Let the summation of $M_i$ be $M$.

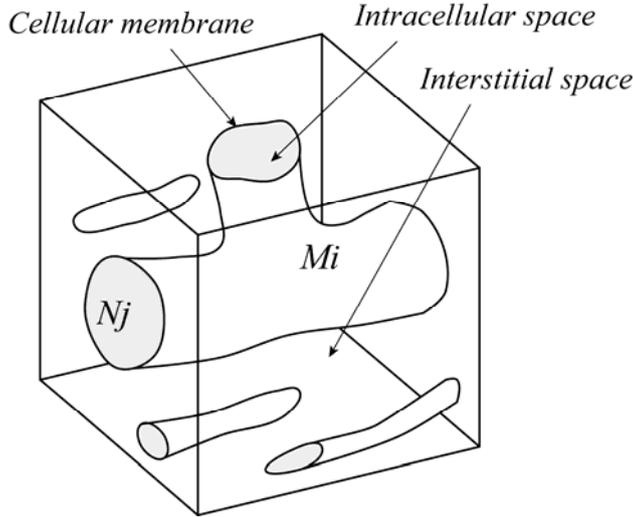

Fig.1 Schematic drawing of a mesoscale volume that includes cell element. A volume $V$ is enclosed by a closed surface $S$. Through $V$, core conductors with surface $M_i$ intersecting $S$ in disk $N_j$ are passing. Let the total summation of $N_j$ be $S'_{int}$, and let the surface $S$, less the disks $S'_{int}$, be $S'_{ext}$. Let the summation of $M_i$ be $M$. Cores have volume $O_i$ within $S$. Let the total summation of $O_i$ be $V'_{int}$, and let the volume $V$, less the volume $V'_{int}$ be $V'_{ext}$.

If we apply Gauss's theorem to the total current $J_{total}$, in the volume $V$ and surface $S$, we obtain the following.

$$\iiint_V \nabla \bullet J_{total} \, dv = \iint_S J_{total} \cdot n \, ds = 0 \qquad (45)$$

Equation (16) states that the term inside the integral on the left side of the eq. (45) is zero, and hence this integral is zero. Therefore the right-hand side, the summation of the total current $J_{total}$ over the closed surface $S$ containing volume $V$, is zero. The same can be said to the total current $J_{total}$, in the volume $V'_{ext}$.

$$\iiint_{V'_{ext}} \nabla \bullet J_{total} \, dv = \iint_{S'_{ext}} J_{total} \bullet n \, da + \iint_M J_{total} \bullet n \, da = 0 \qquad (46)$$

Note that the surface $S'_{ext}$ is not a closed surface because $S'_{int}$ is excluded from the surface $S$. The surface $S'_{ext} + M$ is the closed surface enclosing the volume $V'_{ext}$. Similarly to the eq. (45), the left-hand side of eq. (46) is zero. The right-hand side of eq. (46) implies that the "summation of the total current $J_{total}$ over the boundary that connect the extracellular space to that of the neighboring volume is equal to the summation of $J_{total}$ over the cell surface contained in the volume $V$ ".

$$\iint_{S'_{ext}} J_{total} \bullet n\, da = -\iint_M J_{total} \bullet n\, da \qquad (47)$$

This relation is the basis for deriving the equation for the current source density. From the left-hand side of eq. (47), we will derive the apparent extracellular current, and from the right-hand side, we will derive the current source density.

## 3-2. Apparent extracellular current and apparent extracellular space

*Apparent extracellular current*

**Defining a microscopic current density**: Consider a microscopic imaginary current density $J^a(x,t)$ defined for all location $x(x,y,z)$ in the volume $V$. Let the $\nabla \bullet J^a$ at any location within the volume $V$ takes the same value $<\nabla \bullet J^a>$, and assign value for $<\nabla \bullet J^a>$ by equalizing the integration of $<\nabla \bullet J^a>$ over the volume $V$ with the summation of the total current $J_{total}$ over the surface $S'_{ext}$. Because the surface $S'_{ext}$ is not a closed surface, $<\nabla \bullet J^a>$ is not the average of $\nabla \bullet J_{total}$ and may have non-zero value.

$$\iint_{S'_{ext}} J_{total} \bullet n\, da = \iiint_V <\nabla \bullet J^a(x,t)> dv = <\nabla \bullet J^a(x,t)> \iiint_V dv = <\nabla \bullet J^a(x,t)> V \qquad (48)$$

$$<\nabla \bullet J^a(x,t)> = \frac{1}{V} \iint_{S'_{ext}} J_{total} \bullet n\, da \qquad (49)$$

**Coarse graining**: Here we introduce a mesoscopic scale 3D coordinate system $X(X,Y,Z)$ in which the volume $V$ converge to an infinitesimally volume element $\delta V$. Define a novel current density $J^A(X,t)$ for any location $X(X,Y,Z)$ by equalizing $\nabla \bullet J^A$, the divergence of $J^A(X,t)$, with the integration of $<\nabla \bullet J^a>$ over the volume $V$, i.e. the integration of total current $J_{total}$ over the surface $S'_{ext}$. With this current density, the left-hand side of eq. (47) can be written as follows.

$$\iint_{S'_{ext}} J_{total} \bullet n\, da = <\nabla \bullet J^a(x,t)> V \Rightarrow \nabla \bullet J^A(X,t)\delta V \qquad (50)$$

Here the value of $\nabla \bullet J^A(X,t)$ is as follows.

$$\nabla \bullet J^A(X,t) = <\nabla \bullet J^A(x,t)> \frac{V}{\delta V} = \frac{1}{\delta V} \iint_{S'_{ext}} J_{total} \bullet n\, da \qquad (51)$$

In this evaluation, the divergence should be calculated with respect to the mesoscopic coordinate system $X(X,Y,Z)$. We need to define the three components of $J^A(X,t)$. To do so, we can assume that the volume $V$ is a cube with its all surfaces align to the axis, and name the sum of the pair of two parallel surfaces of $S'_{ext}$ perpendicular to $x$, $y$ and $z$ as $S'_{ext,x}$, $S'_{ext,y}$, $S'_{ext,z}$ respectively. With this notation of the surfaces, we can define $J^A(X,t)$ as follows.

$$J^A(X,t) = \begin{pmatrix} J_x^A \\ J_y^A \\ J_z^A \end{pmatrix} = \begin{pmatrix} \iint_{S'_{ext,x}} J_{total} \bullet n\, da \\ \iint_{S'_{ext,y}} J_{total} \bullet n\, da \\ \iint_{S'_{ext,y}} J_{total} \bullet n\, da \end{pmatrix} \Big/ \delta V \qquad (52)$$

We refer to $J^A(X,t)$ as the **apparent extracellular current density**. The current density defined this manner is an imaginary current density, not the genuine extracellular current density.

*Apparent extracellular space*

In the process of coarse graining, the distinction between the intracellular space and the interstitial space is lost. Hence we need to redefine the extracellular potential and the field. We define the average of extracellular potential $\phi_{ext}(x,t)$ over the volume $V$ in the microscopic coordinate as the extracellular potential $\phi_{ext}(X,t)$ in the mesoscopic coordinate. The extracellular field $E(X,t)$ should be related to the potential as follows.

$$E(X,t) = -\nabla \phi_{ext}(X,t) \tag{53}$$

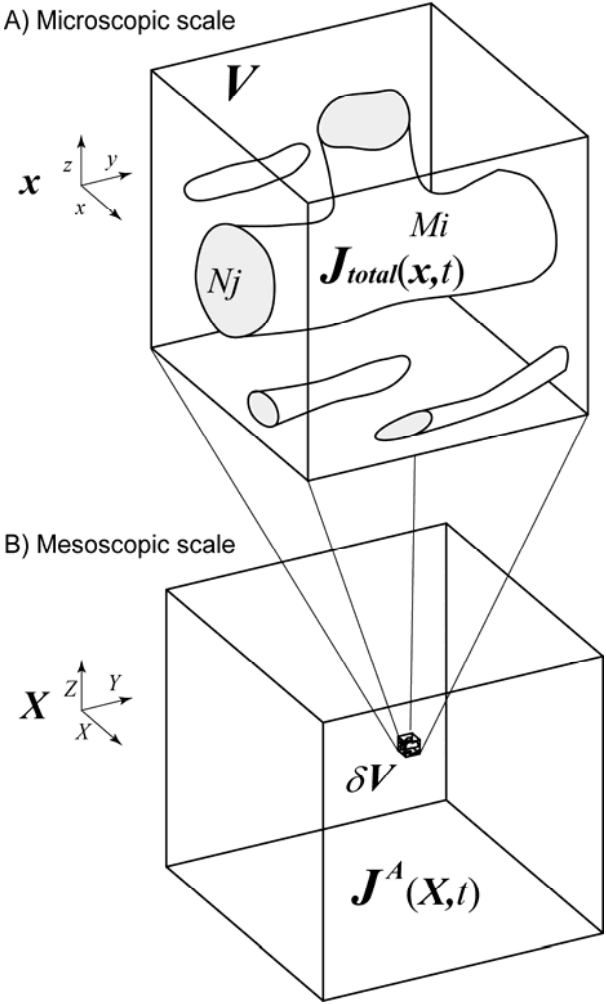

Fig.2 Coarse graining of extracellular space. A) Current density $J_{total}$ is defined for any location in the microscopic coordinate system $x(x,y,z)$. B) Introduce a mesoscopic coordinate system $X(X,Y,Z)$ in which the volume $V$ converge to a small volume $\delta V$. Define a novel current density $J^A(X,t)$ for any location $X(X,Y,Z)$ in this space.

We can assume that the apparent extracellular current $J^A(X,t)$ is composed of the **apparent free current** $J'_{free}$ and the **apparent displacement current** $\partial D'/\partial t$.

$$\boxed{J^A(X,t) = J'_{free}(X,t) + \frac{\partial D'(X,t)}{\partial t}} \tag{54}$$

In biological tissue, we can disregard the convection current $J_{convection}$ and take only the conduction current $J_{conduction}$ into account as the free current. If we assume linear constituent relations for the conduction and displacement current, we can define $D'$ and $J'_{free}$ as the following.

$$D'(X,t) = \int \varepsilon'(X,\tau) E(X,t-\tau) d\tau \tag{55}$$

$$J'_{free}(X,t) = \int \sigma'(X,\tau) E(X,t-\tau) d\tau \tag{56}$$

Here $\sigma'(X,t)$ and $\varepsilon'(X,t)$ are the **apparent conductivity** and **apparent permittivity** respectively defined for the imaginary apparent extracellular space, not the conductivity and permittivity of the fluid that fill the interstitial space. If the conductivity and permittivity are uniform and constant over space and time, then we obtain the following.

$$\boxed{J^A(X,t) = \sigma' E(X,t) + \frac{\partial(\varepsilon' E(X,t))}{\partial t}} \quad (57)$$

The values of $\sigma'$ and $\varepsilon'$ may be quite different from those of the materials in the interstitial space. We will touch upon the mechanisms that determine the apparent conductivity and permittivity in Section 4 of this article.

### 3-3. Current source density as the divergence of apparent extracellular current density

*Membrane current density $I_m$*

The right-hand side of eq. (47) can be coarse grained in the same manner as the left-hand side shown above. Consider a microscopic imaginary surface current density $I_m(x,y,z)$ for all locations on the surface $M$. Let $I_m$ be perpendicular to the surface, and the amplitude takes a uniform value of $<I_m>$ over $M$. With regard to the direction of $I_m$, we choose the direction from within the core toward the volume $V'_{ext}$ across the surface $M$ be positive because it is the convention in electrophysiology.

$$\iint_M J_{total} \bullet n \, da = \iint_M <-I_m(x,t)> da = -<I_m(x,t)> M \quad (58)$$

In the mesoscopic coordinate system $X(X,Y,Z)$, in which the volume $V$ is considered to be a small volume $\delta V$, let surface $M$ become $\delta M$. Define a novel current density $I_m(X,t)$, which we refer to as the **membrane current density**, for any location $X(X,Y,Z)$. With this current density, the right-hand side of eq. (47) can be written as follows.

$$-\iint_M J_{total} \bullet n \, da = <I_m(x,t)> M \Rightarrow I_m(X,t)\delta M \quad (59)$$

By substituting (59) and (50) into eq. (47), we obtain

$$\nabla \bullet J^A(X,t)\delta V = I_m(X,t)\delta M \quad (60)$$

Let $\beta(X)$ be the surface density (or membrane density), specifically $\beta(X) = M/V = \delta M/\delta V$. Then

$$\boxed{\nabla \bullet J^A(X,t) = \beta(X) \cdot I_m(X,t)} \quad (61)$$

Note that $I_m(X,t)$ is a surface current density.

*Current source density $I_{CSD}$*

Instead of $I_m$, we can consider a microscopic imaginary current density $I_{CSD}(x,y,z)$ for all locations in the volume $V$. Let $<I_{CSD}>$ be the summation of $J_{total}$ over the surface $M$ divided by the volume $V$. Then we obtain

$$\iint_M J_{total} \bullet n \, da = \iiint_V <-I_{CSD}(x,t)> dv = -<I_{CSD}(x,t)> V \quad (62)$$

Consider a novel current density $I_{CSD}(X,t)$, which we refer to as the **current source density**, defined for any location $X(X,Y,Z)$ in the mesoscopic coordinate. Then

$$\iint_M J_{total} \bullet n \, da = -<I_{CSD}(x,t)> V \Rightarrow -I_{CSD}(X,t)\delta V \quad (63)$$

By substituting (63) and (50) into eq. (47), we obtain

$$\nabla \bullet J^A(X,t)\delta V = I_{CSD}(X,t)\delta V \quad (64)$$

,and hence
$$\nabla \bullet J^A(X,t) = I_{CSD}(X,t) \quad (65)$$

This eq. (65) is the basis for the current source density analysis. Note that $I_{CSD}(X,t)$ is a volume current density, not a surface current density unlike $I_m(X,t)$. The current density $I_{CSD}(X,t)$ is related to $I_m(X,t)$ as

$$I_{CSD}(X,t) = I_m(X,t)\frac{\delta M}{\delta V} = I_m(X,t)\beta(X) \quad (66)$$

*General expression of the current source density analysis*

By substituting eq. (54) into eq. (65), we obtain

$$\nabla \cdot \left( J_{free}'(x,t) + \frac{\partial D'(x,t)}{\partial t} \right) = I_{CSD}(x,t) \quad (67)$$

Here we replace $X$ with $x$ for convenience. This is the **general equation for the current source density analysis**. At first glance, it may seem to violate the conservation of current because the left-hand side of eq. (67) looks similar to that of the eq. (16), yet the right-hand side of both equations are different. But then eq. (67) was derived from eq. (16). The definitions of $J_{free}'$ and $D'$ make the difference in these equations. Equation (65) may look similar to eq. (17) that states conservation of charge, but they are not. Equation (17) deals with the generation of charge, which is not taken into consideration here.

In practice, one calculates $I_{CSD}(x,t)$ from the measurement of extracellular potential $\phi_{ext}(x,t)$. If the apparent conduction current and displacement current are linearly related to the electric field and the apparent conductivity and apparent permittivity depend on the location and time, we need to use eq. (55) and (56) to describe currents. In such cases, it is not straight forward to calculate the current source density from $\phi_{ext}(x,t)$. Using the Fourier transform, after some calculations, we obtain the following equations, where $\omega$ denotes frequency.

$$(2\pi)\nabla \cdot (\sigma'(x,\omega) + i\omega\varepsilon'(x,\omega))(-\nabla\phi_{ext}(x,\omega)) = I_{CSD}(x,\omega) \quad (68)$$

$$I_{CSD}(x,t) = (2\pi)\int \nabla \cdot (\sigma'(x,\omega) + i\omega\varepsilon'(x,\omega))(-\nabla\phi_{ext}(x,\omega))e^{i\omega t}d\omega \quad (69)$$

These are the equations for the current source density in linear media derived from eq. (67).

If the apparent conductivity and apparent permittivity depend on time but do not depend on the location, then the equation can be written as the following.

$$\nabla \cdot \left( \sigma'(t)E(x,t) + \partial(\varepsilon'(t)E(x,t))/\partial t \right) = I_{CSD}(x,t) \quad (70)$$

In this case, we obtain the following by the Fourier transform.

$$I_{CSD}(x,\omega) = -(2\pi)(\sigma'(\omega) + i\omega\varepsilon'(\omega))(\nabla^2\phi_{ext}(x,\omega)) \quad (71)$$

$$I_{CSD}(x,t) = -(2\pi)\int (\sigma'(\omega) + i\omega\varepsilon'(\omega))(\nabla^2\phi_{ext}(x,\omega))e^{i\omega t}d\omega \quad (72)$$

If the apparent conductivity and apparent permittivity do not depend on the location and the time, then the equations can be written as the following.

$$I_{CSD}(x,\omega) = -(2\pi)(\sigma' + i\omega\varepsilon')(\nabla^2\phi_{ext}(x,\omega)) \quad (73)$$

$$I_{CSD}(x,t) = -(2\pi)\int (\sigma' + i\omega\varepsilon')(\nabla^2\phi_{ext}(x,\omega))e^{i\omega t}d\omega \quad (74)$$

*Conventional current source density analysis*

If the product of permittivity $\varepsilon'$ with the frequency is sufficiently small compared to the conductivity $\sigma'$, i.e. $\omega \varepsilon'/\sigma' \ll 1$, and the conductivity $\sigma'$ does not depend on time, then eq. (69) becomes quite simple.

$$\nabla \cdot (\sigma' \nabla \phi_{ext}(x,t)) = -I_{CSD}(x,t) \tag{75}$$

Further, if $\sigma'$ does not depend on location and isotropic, we obtain the following.

$$\boxed{\sigma' \nabla^2 \phi_{ext}(x,t) = -I_{CSD}(x,t)} \tag{76}$$

This is the **equation for conventional current source density analysis**. The form of this equation is identical to Poisson's equation in electrodynamics. Poisson's equation, however, is obtained by substituting field $E$ into the relation $E = -\nabla \phi$ in Gauss's law, which states the continuity of electric flux. By contrast, eq. (67) above describes relation between the extracellular current and the current source density, stating the continuity of apparent extracellular current. Thus these equations have different physical meanings.

*Monopole issue*

There has been an issue whether current monopoles emerge in the brain (Bedard & Destexhe, 2011; Riera et al., 2012). In the conventional current source density analysis, the integration of $I_{CSD}$ over the region of interest at certain point in time sometimes exhibits significant non-zero value suggesting that trans-membrane current may give rise to current monopoles. Within the framework of the discussion in this article, however, we can say that current monopoles would not emerge if one integrates over a sufficiently large volume to include all parts of the cells or structures involved.

As evident from eq. (46), (59) and (62), integrating $I_{CSD}(X,t)$ or $I_m(X,t)$ over a volume is equivalent to integrate $J_{total}(x,t)$ over the total core surfaces contained in the volume. In case of eq. (46), we assume that cores are not necessarily contained in the volume, and hence the surface $M$ is not closed. This is exactly why the integrals may have non-zero values. But, if we consider integrating over a sufficiently large volume, then all the cores will be entirely contained in the volume, and the summation of the core surfaces will be closed. Continuity of the flux $J_{total}$ ensures that the integration of $J_{total}$ over the area of a closed surface will be zero. The argument in this article assumes that the quasi-static condition, i.e. $\nabla \times E = 0$, holds. Within the quasi-static regime, the argument above is valid whether the media is linear or non-linear. Hence, we can say that monopole would not emerge under quasi-static conditions.

If one integrates $I_{CSD}(X,t)$ or $I_m(X,t)$ over some volume and obtains a non-zero value, it could be due to one or combinations of the following; 1) the area for the integration was not sufficiently large to contain all the structures involved, 2) the contribution of the capacitive component of current in the extracellular space was not properly taken into consideration, 3) the quasi-static condition was not satisfied in the tissue in question.

# 4. Intuitive Account for the Apparent Extracellular Space

4-1. Dielectric dispersion in biological tissue

In this article, we introduced peculiar quantities, the apparent permittivity $\varepsilon'$ and apparent conductivity $\sigma'$. What exactly are these? It is a common practice to assume the brain tissue to be purely ohmic and ignore the capacitive current components in estimating the source for EEG and LFP signals. However, the permittivity of most biological tissue is large and conductivity is small in low frequency range. In various dielectric materials, it is known that the permittivity and conductivity depend on the frequency of electric fields, and this phenomenon has been referred to as the **dielectric dispersion** (Fig.3).   In case of cell suspension, the relative permittivity is lower than $10^2$ in the frequency range higher than 1GHz, and the value could be higher than $10^6$ in the range lower than 1kHz (Cole & Cole, 1941; Cole, 1968; Foster & Schwan, 1989; Martinsen et al., 2002). According to Gabriel et al. (1996), the relative permittivity of brain tissue at 10Hz is near $10^8$. These values in low frequency range are much higher than those of pure water ($\approx 80$) and lipid ($\approx 2$).

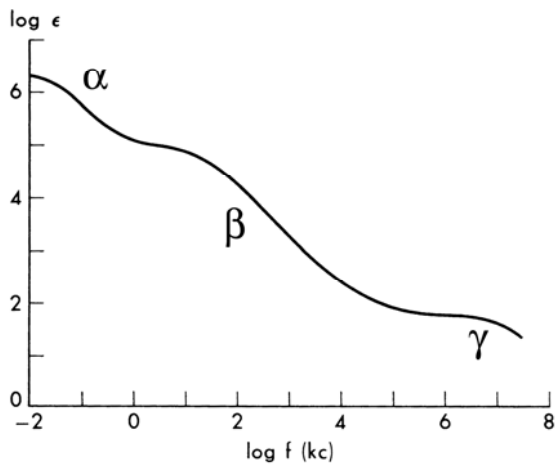

Fig.3 A representation of the principal dispersions of tissue in terms of their relative dielectric constants $\varepsilon$ as a function of frequency. Based on red blood cell suspensions, the β region is explained by the cell membrane capacity. The α dispersion probably represents the relaxation of a surface conductance on these cells or of ion conductance in an excitable membrane. The γ dispersion is mostly the dielectric relaxation of intracellular proteins. (modified from Cole, 1968)

When a sinusoidal electric field of frequency $f$ is applied to a dielectric material, the ratio of the capacitive component of the current to the resistive component should be $2\pi f \varepsilon(f)/\sigma(f)$, where $\varepsilon(f)$ and $\sigma(f)$ are the permittivity and conductivity at the frequency $f$.   By substituting values in Gabriel's study, the ratio reveals a local peak value, approximately 0.9, at 10Hz in addition to the overall tendency of low value in low frequency and high in high frequency range. In case of a simple discrete electric filter circuit with a parallel capacitance and resistance of constant value, the capacitive current can be safely ignored in the low frequency range. The situation, however, is quite different in biological tissue with strong dielectric dispersion. If one takes the measurements by Gabriel seriously, it is not reasonable to ignore capacitive components in the frequency range near 10Hz.

The dielectric dispersion is considered to be due to various electrical polarization processes with various relaxation times. Dispersion in the frequency range higher than 1GHz is due to fast polarization processes, such as electronic polarization and orientation polarization, that have relaxation time ranging from 1psec to 1nsec (Landau et al., 1982; Jackson, 1998; Takashima, 1989). The mechanism for dispersion in the low frequency range is not fully understood. The leading theory referred to as the **interfacial polarization theory** or the **Maxwell-Wagner polarization theory** (Takashima, 1989; Foster & Schwan, 1989; Hanai, 1960) predicts that suspension of dielectric material in a media with different permittivity exhibits large permittivity as the frequency decreases, and the permittivity of the bulk mixture is larger than those of the particles and media. Pauly & Schwan (1959) extended this theory to membrane-covered spheres to show that, in cell suspension in electrolyte, the presence of thin lipid membrane gives rise to dispersion in the frequency range between 1MHz and 1kHz. Another theory called the **counter-ion polarization theory** has been proposed （Foster & Schwan, 1989; Bedard & Destexhe, 2011) for dispersion in frequency range lower than 1kHz. The permittivity and conductivity that appear in the literature discussed above correspond to the apparent permittivity $\varepsilon'$ and conductivity $\sigma'$ defined in the present article. The values for the fluid filling the interstitial space or the intracellular space should be determined by the composition of the electrolytes and biological macromolecules in the fluid. The values for the cell membrane should depend on the composition of the membrane lipid and membrane proteins. The values for the bulk tissue, however, are not some sort of average of the values for the materials included in the tissue. They depend strongly on the morphology of cells and the composition of the tissue, in the low frequency range in particular. (We have shown recently that the presence of long cylindrical neurites gives rise to strong dispersion in the frequency range between 1Hz to 1kHz (Monai et al., submitted).)

| Animal | Region | $\alpha$ | $\lambda$ | Ref. |
|---|---|---|---|---|
| Rat | Cerebellum, molecular layer | $0.21^a$ | $1.55^a$ | Nicholson and Phillips, 1981 |
| | Neocortex | 0.18 | 1.57 | Cserr et al., 1991 |
| | Neocortex | 0.18 | 1.40 | Lundbæk and Hansen, 1992 |
| | Neocortex | 0.20–0.22 | 1.63 | Lehmenkühler et al., 1993 |
| | Hippocampus, CA1 | 0.12 | 1.67 | McBain et al., 1986 |
| | Hippocampus, CA3 | 0.18 | 1.83 | McBain et al., 1986 |
| | Neostriatum | 0.21 | 1.54 | Rice and Nicholson, 1991 |
| | Spinal cord, dorsal horn | 0.20 | 1.62 | Syková et al., 1994 |
| Guinea pig | Cerebellum, molecular layer | $0.28^a$ | $1.84^a$ | Hounsgaard and Nicholson, 1983 |
| Turtle | Cerebellum, molecular layer | 0.31 | 1.44, 1.95, 1.58 | Rice et al., 1993 |
| | Cerebellum, granular layer | 0.21 | 1.77 | Rice et al., 1993 |
| Skate | Cerebellum, molecular layer | $0.24^a$ | $1.62^a$ | Nicholson and Rice, 1986 |
| Cuttlefish | Optical lobe | 0.29 | 1.86 | Nicholson et al., 1995 |
| | Vertical lobe | 0.10 | 1.65 | Nicholson et al., 1995 |

$^a$These cerebellar measurements in the molecular layer did not take anisotropy into account.

Table.1 Values of volume fraction ($\alpha$) and tortuosity ($\lambda$) obtained with the TMA-method (from Nicholson, 1995)

4-2. Apparent extracellular space with apparent permittivity and conductivity

When one studies the diffusion of chemicals such as neurotransmitters, the term "extracellular space" would be used as a synonym for interstitial space (e.g. Nicholson et al., 2000). Many studies reported that the fraction of interstitial space in brain tissue is 10 to 30 % of the tissue volume (Table 1). The term "extracellular space", however, means something different when one studies the spread of electric currents in biological tissue. It is an abstract concept; an imaginary apparent space in which current either generated by cells or provided by external mechanisms flows.

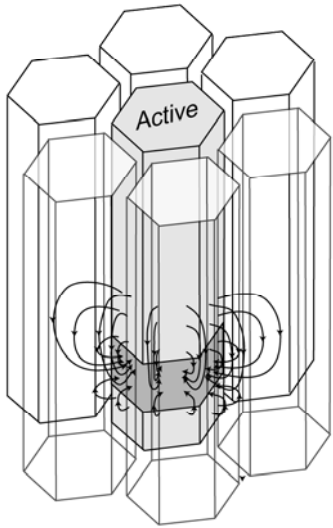
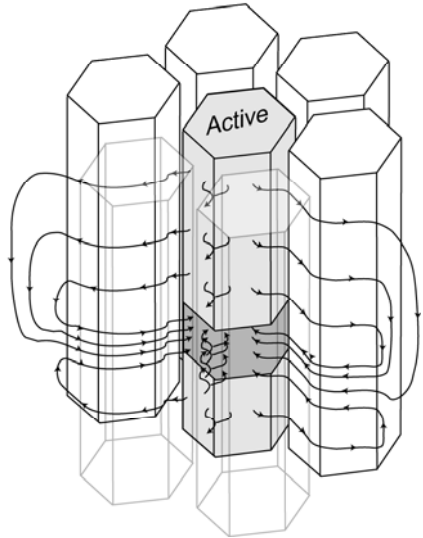

Fig.4 Flow of current generated at a restricted area (dark grey) of a neurite (light gray) as inward transmembrane current.
A) High frequency current flow across the membrane and intracellular space of nearby cells. B) Low frequency current flow mainly in the interstitial space.

The entity of the extracellular space depends on the frequency of the current (Fig.4). In case of a current with very high frequency, the current can flow across the cell membranes as capacitive current. Therefore the extracellular space in this situation would be the entire space including the interstitial space, intracellular space and the membrane (Fig.4A). Hence the density of current in the extracellular space decays quickly with the distance from the current source. In case of a low frequency current, current would not easily flow across the membrane. In this situation, the interstitial space mainly constitutes the extracellular space, and hence the density of the extracellular current only slowly decays with the distance (Fig.4B). Other than these extreme situations, the extracellular space in electrical sense implies an imaginary apparent space composed of the contributions due to the presence of membranes of nearby cells and the intracellular space in addition to the interstitial space. The electrical properties of the **apparent extracellular space** as dielectrics can be described by the **apparent permittivity and conductance** as have been explained in the previous section. They are different from those of the interstitial space, intracellular space and the membrane. They are not the simple average of them.

On the basis of a simple model calculation, Attwell (2003) argued that most of the current applied to biological tissue would flow within the interstitial space and the amount of the displacement current would be negligible in the extracellular space. A tacit assumption in his argument was that if the capacitive current across the membrane is negligibly small, then the displacement current of the extracellular space can be ignored. His calculation shows that the amount of current that flow across the cellular membrane become significant only at frequencies above 700 kHz. But, as has been explained above, the apparent permittivity in the extracellular space is not a simple reflection of the membrane capacitance, nor does the capacitive current in the apparent extracellular space reflect the current flow through the capacitance of the membrane.

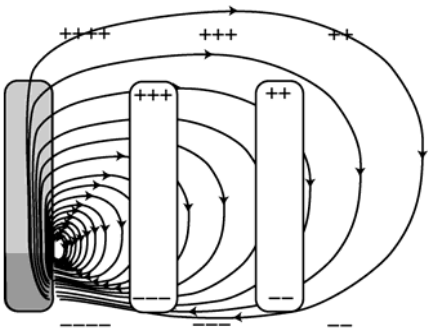

A) Primary current

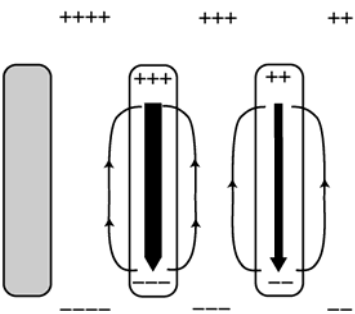

B) Secondary current

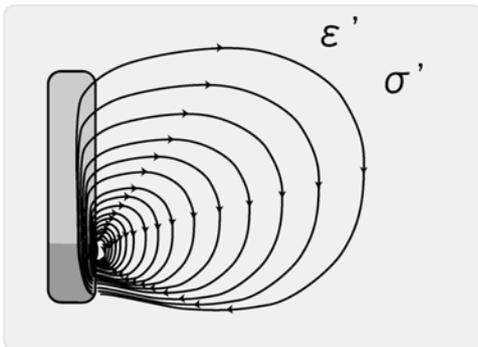

C) Overall current

Fig.5 Apparent permittivity and conductivity in the extracellular space. Consider cylindrical neurons aligned parallel. Assume that the membrane have constant frequency independent permittivity and conductivity. Assume that the permittivity of the interstitial and intracellular space are negligibly small. A) Let one of the neuron generates ionic membrane current at one point along the cylinder and capacitive return current somewhere else providing primary sink and source. Most of the primary current flows within the interstitial space and gives rise to change in the potential gradient in the interstitial space. B) The intracellular space in the nearby cylinders has the same potential gradient as the interstitial space. This gradient gives rise to intracellular current along the longitudinal axis of the cylinders, and the current flows across the membrane of the cylinder to complete a loop providing a secondary sink and source in the interstitial space. This secondary current flows in the direction opposite to the primary current and gives rise to the secondary change in the potential in the interstitial space. C) The overall current flow and the potential in the interstitial space can be related to the initial membrane current with the apparent permittivity $\varepsilon'$ and conductivity $\sigma'$ of the apparent extracellular space.

In order to understand the concept of the apparent extracellular space qualitatively, let us consider a situation in which cylindrical neurons are aligned parallel (Fig.5). Here we assume that the membrane have frequency-independent permittivity and conductivity. For the sake of simplicity, let us assume that the permittivity of the interstitial and intracellular space is negligibly small. Now let one of the neuron generates an inward ionic membrane current at a restricted area along the cylinder and the capacitive return current elsewhere providing primary sink and source respectively to the interstitial space. As argued by Attwell, most of the primary current flows within the interstitial space and gives rise to change in the potential gradient in the interstitial space (Fig.5A). Before this change in potential gradient becomes significant, only negligible amount of current flows across the membrane of nearby cylinders. When the potential gradient, i.e. the electric field, in the interstitial space is generated, there is the same potential gradient in the intracellular space within the nearby cylinders as well. This gradient then gives rise to the intracellular current along the longitudinal axis of the cylinders. The current should flow across the membrane of the cylinder to complete a current loop providing a secondary sink and source in the interstitial space. This secondary current flows in the direction opposite to the primary current and gives rise to secondary change in the potential within the interstitial space (Fig.5B). To relate the initial membrane current with the final potential in the interstitial space, we use the concept of the apparent extracellular space with apparent permittivity and conductivity (Fig.5C). The intracellular current due to the potential gradient generated in this sequence of events can be considered to be a part of the relaxation processes responsible for the slow dielectric polarization of the apparent extracellular space. The relaxation time constant is large because of the large capacitance and the resistance of the membrane, hence the dielectric dispersion in low frequency range.

## Acknowledgements

We thank Jorge J. Riera, Katayama Mitunori and Asami Koji for helpful discussions and comments. H.M. thanks William Ross for providing H.M. with time, space and atmosphere at the early stage of this work.